\documentclass[12pt]{article}
\usepackage{amssymb}
\usepackage{amsmath,amsfonts,feynmp}
\usepackage{tipa}
\usepackage{stmaryrd}
\usepackage{graphicx}
\usepackage{color}

\makeatletter\@addtoreset{equation}{section}\makeatother

\def\be{\begin{equation}}
\def\ee{\end{equation}}
\def\bea{\begin{eqnarray}}
\def\eea{\end{eqnarray}}

\makeatletter\@addtoreset{equation}{section}\makeatother

\renewcommand{\title}[1]{\vbox{\center\LARGE{#1}}\vspace{5mm}}
\renewcommand{\author}[1]{\vbox{\center#1}\vspace{5mm}}
\newcommand{\address}[1]{\vbox{\center\em#1}}

\begin{document}

\unitlength = .8mm

\begin{titlepage}
\begin{center}
\hfill \\
\hfill \\
\vskip 1cm

\title{ Holographic Superconductors from Einstein-Maxwell-Dilaton Gravity }

{Yan Liu \footnote{Email: liuyan@itp.ac.cn}} and {Ya-Wen Sun
 \footnote{Email: sunyw@itp.ac.cn}}

\address{ Key Laboratory of Frontiers in Theoretical Physics
\\ Institute of Theoretical Physics, Chinese Academy of Sciences,
\\P.O. Box 2735, Beijing 100190, China}

\end{center}

\vskip 1cm

\abstract{We construct holographic superconductors from Einstein-Maxwell-dilaton gravity in 3+1 dimensions with two adjustable couplings $\alpha$ and the charge $q$ carried by the scalar field.
For the values of $\alpha$ and $q$ we consider, there is always a critical temperature at which a second order phase transition occurs between a hairy black hole and the AdS RN black hole in the canonical ensemble, which can be identified with the superconducting phase transition of the dual field theory. We calculate the electric conductivity of the dual superconductor and find that for the values of $\alpha$ and $q$ where $\alpha/q$ is small the dual superconductor has similar properties to the minimal model, while for the values of $\alpha$ and $q$ where $\alpha/q$ is large enough, the electric conductivity of the dual superconductor exhibits novel properties at low frequencies where it shows a ``Drude Peak" in the real part of the conductivity.}

\vfill

\end{titlepage}

%%%\eject \tableofcontents%

\section{Introduction}
 The AdS/CFT correspondence \cite{Aharony:1999ti} provides an
 elegant idea to the study of strongly coupled
quantum field theories by relating them to certain classical gravity theories or string systems. Nowadays, the holographic correspondence has also become a very efficient method to deal with the strongly interacting systems in condensed matter physics. A great deal of progress has been made in the application of this holographic method to condensed matter physics. Some nice reviews on this subject can be found in \cite{Hartnoll:2009sz,{Herzog:2009xv},
{McGreevy:2009xe},{Horowitz:2010gk},{Sachdev:2010ch},{Kaminski:2010zu}}.

The high temperature superconductor is an exciting while not
completely understood subject in condensed matter physics
and it remains an unsolved mystery because its theoretical
basis might be a strongly coupled field theory. It is
interesting to use the gauge/gravity duality to obtain
some insights into the properties of superconductors. The simplest
model to obtain a holographic superconductor with quite similar behavior to real superconductors was first built in \cite{{Gubser:2008px},Hartnoll:2008vx} through
Einstein gravity which is minimally coupled to a Maxwell
field and a charged complex scalar with a potential
term. Below some critical temperature $T_c$, the charged black hole solutions
develop a non-trivial hair. From the point of view of the dual field theory a $U(1)$ symmetry breaks below $T_c$ at a finite charged density because of the condensation of a charged scalar.\footnote{  Note that, according to the dictionary of AdS/CFT, this $U(1)$ symmetry on the field theory side should be a global one, thus the dual field theory is superfluid \cite{{Basu:2008st},Herzog:2008he}. Discussions about how to make the symmetry a local one can be found in \cite{Domenech:2010nf,{Maeda:2010br}}. We assume that this $U(1)$ symmetry will eventually be gauged.} This model naturally realized s-wave superconductors and in this paper we will call this model the minimal model for simplicity. Complete analysis including the backreactions of this system and about the zero temperature limit has been considered in \cite{Hartnoll:2008kx,{Gubser:2009cg},{Horowitz:2009ij},{Konoplya:2009hv}}. Later, following \cite{Liu:2009dm,{Cubrovic:2009ye},{Faulkner:2009wj}}, the fermion spectral function in this system was analyzed and very similar behavior to what was seen in the angle-resolved photoemission experiments (ARPES) on high $T_c$ cuprates was found in \cite{Chen:2009pt,{Faulkner:2009am},{Gubser:2009dt}}. The realization of p-wave superconductors has been studied in \cite{{Gubser:2008wv},Roberts:2008ns},
and d-wave superconductors in \cite{Chen:2010mk,{Herzog:2010vz},{Benini:2010qc}}.

Although the minimal holographic superconductor model has
achieved much success, it is still necessary to consider
 more generalized non-minimal holographic superconductors to incorporate more (or some universal) features of superconductors in real physical systems, or even to find the dual description of real superconductors.
 In \cite{{Franco:2009yz}, Aprile:2009ai}, a general class of
 superconductors was considered and some universal behavior was found. Discussions on other
aspects of generalized holographic superconductors can be
found {\it e.g.} in \cite{Aprile:2010yb,{Pan:2010at}}.

One of these interesting generalized holographic superconductor
models \cite{Aprile:2009ai} is the Einstein-Maxwell-dilaton model.\footnote{The terminology ``dilaton" is not so accurate here. Generally ``dilaton" refers to a real scalar field
which is non-minimally coupled to the Einstein-Maxwell theory. Hairy black hole solutions widely exist in
Einstein-Maxwell-dilaton
theory and the holographic dual for Einstein-Maxwell-dilaton model
is itself
very interesting because Einstein-Maxwell-dilaton gravity is very common in the low-energy effective
theories of string theory
and black hole solutions in Einstein-Maxwell-dilaton gravity may
exhibit some
quite special thermal properties \cite{Goldstein:2009cv}. The holography
of dilaton black holes has been
studied in
\cite{{Herzog:2009gd}, Gubser:2009qt,{Goldstein:2009cv},{Cadoni:2009xm}} (see also
\cite{{Charmousis:2010zz},{Perlmutter:2010qu},{Taylor:2008tg},{Chen:2010kn},
{Cai:1996eg},{Cai:1997},{Charmousis:2009},{lee:1006}}). However, in the framework of holographic superconductors the scalar field arising in the bulk theory has to be a complex one as it has to be charged under the Maxwell field, and in the title and the remainder of this paper by ``Einstein-Maxwell-dilaton model" we actually mean the Einstein-Maxwell gravity non-minimally coupled to a complex scalar in a similar way to the dilaton field.} For our aim to realize holographic superconductors from Einstein-Maxwell-dilaton gravity, there are some additional constraints on the form of the action of the gravity theory. One of these constraints is that AdS Reissner-Nordstr\"{o}m (RN) black holes should be solutions to the theory at a finite charge density. This excludes the $e^{\alpha\eta}F^2$ type Einstein-Maxwell-dilaton models, as the equation of motion for the scalar field $\eta$ in this theory ensures that all
the charged black holes carry non-trivial hair. This motivates people to consider the ${\mathrm{Cosh}}(\alpha\eta)F^2$ type Einstein-Maxwell-dilaton models
where AdS RN black hole solutions can exist. This was first studied in the framework of holographic superconductors in \cite{Aprile:2009ai}, where it was found that this model shares essentially the same
physics as the minimal model in \cite{Hartnoll:2008kx} when $\alpha=1$. In a nice early paper \cite{Cadoni:2009xm},
the phase transition between AdS RN black holes and dilatonic
black holes with neutral dilaton was studied in this type of models and some novel behavior in the electric conductivity at low frequencies was found. However, because the dilaton considered there is neutral, it does not have a dual superconductor description. In this paper, we will consider this model with general values of $\alpha$ and charged dilatons in the framework of holographic superconductors.

For the values of $\alpha$ and charge $q$ we consider, there is always a critical temperature at which a second order thermal phase transition occurs between a hairy black hole and the AdS RN black hole in the canonical ensemble. Below this temperature the dual theory is in a superconducting phase while above this temperature the dual theory is in a normal phase. We also study the electric conductivity of the dual superconductor. For the values of $\alpha$ and the charge $q$ where $\alpha/q$ is small the dual superconductor has similar properties to the minimal model as pointed out in \cite{Aprile:2009ai}. However, for the values of $\alpha$ and $q$ where $\alpha/q$ is large enough, the electric conductivity of the dual superconductor exhibits novel properties which are very different from the minimal model at low frequencies, {\it e.g.} near $\omega\to 0$, a ``Drude Peak" arises. This can also be seen from the shape of the Schr\"{o}dinger potential after translating the calculation of the electric conductivity into a one dimensional scattering problem.

In the remainder of this paper, we will first construct the basics of Einstein-Maxwell-dilaton model in Sec.2. In Sec.3 we give the numerical results of the phase transition between hairy black holes and AdS RN black holes in a canonical ensemble. In Sec.4 we show the behavior of the electric conductivity of the dual superconductor. Sec.5 is devoted to conclusions and discussions.

\section{Basic Set-up for Einstein-Maxwell-Dilaton Models}

In this section we follow \cite{Aprile:2009ai} to consider a generalized holographic
superconductor model built from the most generalized covariant gravity
Lagrangian with at most two derivatives of fields in 3+1 dimensions.
The model has the following field contents: a metric field
$g_{\mu\nu}$, a $U(1)$ gauge field $A_{\mu}$, a real scalar field
$\eta$ and a St\"{u}ckelberg field $\theta$, which are coupled in
the following way
 \be
I=\frac{1}{16\pi G}\int
d^4x\sqrt{-g}\bigg[R-\frac{1}{4}G(\eta)F^{\mu\nu}F_{\mu\nu}+\frac{6}{\ell^2}U(\eta)-\frac{1}{2}(\partial
\eta)^2-\frac{1}{2}J(\eta)(\partial_\mu\theta-A_\mu)^2\bigg],\ee
where $G(\eta)$, $U(\eta)$ and $J(\eta)$ are three functions of the
scalar $\eta$, whose forms can affect the dynamics of the dual
superconductor.

This system has a $U(1)$ gauge symmetry and the gauge
transformations are the standard one $A_{\mu}\to
A_{\mu}+\partial_{\mu}\Lambda, ~\theta\to \theta+\Lambda$, so we can
choose the gauge $\theta=0$ in the following calculations. In fact,
we can interpret the scalar field $\eta$ as the modulus of a complex
scalar $\psi$ and $\theta$ as its phase, {\it i.e.} $\psi=\eta
e^{iq\theta}$.

The Einstein equation of motion for $g_{\mu\nu}$ is
\bea\label{eomgmunu}
R_{\mu\nu}-\frac{1}{2}g_{\mu\nu}\bigg(R+\frac{6}{\ell^2}U(\eta)
-\frac{1}{2}(\partial_\alpha\eta\partial^\alpha\eta)-\frac{1}{4}G(\eta)F^2
-\frac{1}{2}J(\eta)A^2\bigg)&&\nonumber\\
-\frac{1}{2}\partial_\mu\eta\partial_\nu\eta
-\frac{1}{2}J(\eta)A_\mu A_\nu
-\frac{1}{2}G(\eta)F_{\mu\rho}F_{\nu}^{~\rho}=0.&& \eea The equation
of motion for the gauge field $A_{\mu}$ is \be\label{eomamu}
\nabla_\mu \bigg(G(\eta) F^{\mu\nu}\bigg)-J(\eta)A^{\nu}=0,\ee and
the equation of motion for the scalar field $\eta$ is
\be\label{eometa} \nabla_\mu\nabla^\mu\eta-\frac{1}{4}\frac{\partial
G(\eta)}{\partial\eta}F_{\mu\nu}F^{\mu\nu}+\frac{6}{\ell^2}\frac{\partial
U(\eta)}{\partial\eta}-\frac{1}{2}\frac{\partial
J(\eta)}{\partial\eta}A_\mu A^{\mu}=0.\ee

The choices of the three functions $G(\eta)$, $U(\eta)$ and
$J(\eta)$ are crucial to the building of the dual superconductor
here. To build a superconductor from AdS/CFT, we need to have an AdS
vacuum solution in this system, which requires that $U(\eta)$ has a
finite and positive extremum at $\eta=0$. Also we require that AdS
RN black hole is a solution to this system. As a very simple
example, in this paper we consider the following choices of the
functions $G(\eta)$, $U(\eta)$ and $J(\eta)$ as
\bea\label{choice} G(\eta)&=&\mathrm{cosh}(\alpha\eta), \nonumber\\
U(\eta)&=&1-\frac{\ell^2}{12}m^2{\eta^2}, \nonumber\\
J(\eta)&=& q^2\eta^2, \eea where $\alpha$, $m$ and $q$ are
constants.

 Under this choice, the system
has an extra $Z_2$ symmetry: $\eta\to-\eta$.
Note that the AdS RN black hole would not be a solution
to this system if we choose $G(\eta)$ to be of the form
$e^{\alpha\eta}$, which can be easily seen from the equations of
motion. This choice of the three functions (\ref{choice}) has been
studied in \cite{Aprile:2009ai} and \cite{Cadoni:2009xm} for the
cases $\alpha=1$, $q=3$ and $q=0$ with general $\alpha$,
respectively. In \cite{Aprile:2009ai} it was pointed out that for $\alpha=1$ and $q=3$, this model gives a dual superconductor which has very similar properties to the minimal model studied in \cite{Hartnoll:2008kx}. In \cite{Cadoni:2009xm}, the authors found that at $q=0$ there are also phase transitions for general $\alpha$ and as $\alpha$ increases, some novel properties arise. However, as $q=0$ in this model, the $U(1)$ symmetry is not broken and the dual field theory does not have a superconductor description. In this paper we will analyze this model in detail for
general values of $\alpha$ and $q\neq 0$ and find that as the value of $\alpha$ increases the dual superconductor has some different behavior compared to the minimal model discussed in \cite{Hartnoll:2008kx}.

To give the dual superconducting phase, we need a hairy solution
with the form assumed to be
 \bea\label{ansatzsol}
ds^2&=&-g(r)e^{-\chi(r)}dt^2+\frac{dr^2}{g(r)}+r^2(dx^2+dy^2),\nonumber\\
A&=&\phi(r)dt,\nonumber\\
\eta&=&\eta(r).\eea

The equations of motion can be simplified to be \bea
\label{gttgrr1}\chi'+\frac{r}{2}\eta'^{2}+\frac{r}{2g^2}e^{\chi}J(\eta)\phi^2&=&0,\\
\label{gttgrr2}\frac{1}{4}\eta'^{2}+\frac{G(\eta)}{4g}e^{\chi}\phi'^{2}+\frac{g'}{rg}+\frac{1}{r^2}
-\frac{3}{\ell^2g}U(\eta)+\frac{1}{4g^2}e^{\chi}J(\eta)\phi^2&=&0,\\
\label{ax}\phi''+\phi'\bigg(\frac{2}{r}+\frac{\chi'}{2}+\frac{\partial_{\eta} G\eta'}{G}\bigg)
-\frac{J(\eta)}{g G(\eta)}\phi&=&0,\\
\label{eta}\eta''+\eta'\bigg(\frac{2}{r}-\frac{\chi'}{2}+\frac{g'}{g}\bigg)+\frac{1}{2g}e^{\chi}
\partial_\eta G\phi'^{2}+\frac{6}{\ell^2g}\partial_\eta U+\frac{1}{2g^2}e^{\chi}\partial_\eta J\phi^2&=&0, \eea
under the assumption (\ref{ansatzsol}). Note that (\ref{gttgrr1})
and (\ref{gttgrr2}) are the combinations of the Einstein equations
of motion for $g_{tt}$ and $g_{rr}$, while (\ref{ax}) and
(\ref{eta}) are the equations of motion for $A_t$ and $\eta$,
respectively. The equation of motion for $g_{xx}$ is not independent
and it can be derived from
the four equations above.\footnote{Since this is not an obvious
observation, we will give a simple proof in the appendix.}
 We can see that the AdS RN black hole is a solution to this
system with \be\label{rn}\chi(r)=\eta(r)=0,~~~g(r)=\frac{r^2}{\ell^2}-\frac{1}{r}
(\frac{r_+^3}{\ell^2}+\frac{\rho^2}{4r_+}) +\frac{\rho^2}{4r^2} ~~\mathrm{and}~~
\phi=\rho(\frac1{r_+}-\frac1r).\ee

It is difficult to find solutions to the equations of motion
analytically and we will do this in this paper using numerical
methods. We can solve the equations by integrating the fields from the
horizon $r_+$, which is determined by $g(r_+)=0$, to infinity numerically. There are totally four  physical fields which need to be solved: $\eta(r)$, $\phi(r)$, $\chi(r)$ and $g(r)$. We demand that $\phi(r)$ vanish at the horizon in order for the gauge one-form to be well defined at the horizon \cite{Gubser:2008px}. At the horizon there are four independent parameters \be
r_+,~~\eta_+\equiv\eta(r_+),~~E_+=\phi'(r_+),~~\chi_+=\chi(r_+),\ee
as $g(r_+)=0$ and $\eta'(r_+)$ can be determined from the four parameters above using the equations of motion expanded near the horizon:
\bea
\big[rge^{-\chi/2}\big]'-\frac{3r^2}{\ell^2}e^{-\chi/2}U(\eta)+
\frac{r^2G(\eta)}{4}e^{\chi/2}\phi'^{2}&=&0,\nonumber\\
\eta'(r_+)g'(r_+)+\frac{1}{2}e^{\chi_+}\partial_\eta G(\eta_+)E_+^2+\frac{6}{\ell^2}\partial_\eta U(\eta_+)&=&0.
\eea

We can get solutions of the system by integrating the equations of motion given the initial values of the four parameters above at the horizon.

% From (\ref{gttgrr1}) and
%(\ref{gttgrr2}), we obtain \be
%\big[rge^{-\chi/2}\big]'=\frac{3r^2}{\ell^2}e^{-\chi/2}U(\eta)-\frac{r^2G(\eta)}{4}e^{\chi/2}\phi'^{2}.\ee

%In order for the scalar potential has finite potential, we
%should have $\phi(r_+)=0$ while $\phi'(r_+)$ is finite.
%$g'(r_+)$ and $\chi'(r_+)$ could be determined by $\chi(r_+)$,
%$\eta(r_+)$ and $\phi'(r_+)$ with the following relations:

The Hawking temperature for the solution can be calculated as
\be T=\frac{r_+}{16\pi\ell^2}\bigg(12e^{-\chi_+/2}U(\eta_+)-e^{\chi_+/2}
G(\eta_+)E_+^2\ell^2\bigg).\ee

Before doing the numerical calculations, we list the three scaling symmetries of this system, which can help simplify the calculation. The first one is
\be e^{\chi}\to b^2 e^{\chi},~~ t\to bt,~~\phi\to \phi/b,\ee
and we can use this scaling symmetry to set $\chi(r)=0$ at the boundary. The second one is
\be r\to br,~~(t,x,y)\to (t,x,y)/b,~~g\to b^2 g,~~\phi\to b\phi,\ee
which can be used to set $r_+=1$.
The third scaling symmetry is
\be r\to br,~~t\to bt,~~\ell\to b\ell,~~q\to q/b,\ee
which rescales the metric to $b^2g(r)$ and $A=\phi(r)dt$ to $bA$. This scaling symmetry can be used to set $\ell=1$ during the calculations.

At the AdS boundary $r\to\infty$ the behavior of the fields are the following. For the scalar field
\be \eta(r)\sim \frac{\psi^{({\triangle_-})}}{r^{\triangle_-}}+
\frac{\psi^{({\triangle_+})}}{r^{\triangle_+}},\ee where \be\triangle_{\pm}=\frac{3\pm \sqrt{9+4m^2\ell^2}}{2}. \ee
To have a stable theory we need to specify a boundary condition either $\psi^{({\triangle_-})}=0$ or $\psi^{({\triangle_+})}=0$. For $-9/4 <m^2\ell^2<-5/4$, either $\psi^{({\triangle_-})}= 0$ or $\psi^{({\triangle_+})}= 0$ can be chosen as the boundary condition.
For $-5/4 \leq m^2\ell^2$, we can only impose the boundary condition $\psi^{({\triangle_-})} =0$. In this paper we will set $m^2\ell^2=-2$ for simplicity and ${\triangle_-}=1$ while ${\triangle_+}=2$ for this value of $m$. Thus with different choices of boundary conditions we can read off the expectation value of a dimension one operator ${\mathcal{O}}_{1}$ or of a dimension two operator ${\mathcal{O}}_{2}$. For the boundary condition $\psi^{(1)}=0$, we can have \be \langle{\mathcal{O}}_2\rangle=\psi^{(2)},\ee and for the boundary condition
$\psi^{(2)}=0$, we have \be\langle{\mathcal{O}}_1\rangle=\psi^{(1)}.\ee

The boundary behavior of the gauge field is
\be \phi(r)\sim \mu-\frac{\rho}{r},\ee
where $\mu$ is the chemical potential of the dual field theory while $\rho$ is the charge density.
The boundary behavior of the metric fields $\chi(r)$ and $g(r)$ can be determined from the equations of motion to be
\be \chi(r)\sim \frac{\triangle{{\psi}^{(\triangle)}}^2}{4\ell^2}\frac{1}{r^{2\triangle}},\ee
and
\bea
&&g(r)\sim\frac{r^2}{\ell^2}+\frac{\triangle{{\psi}^{(\triangle)}}^2}{4\ell^2}
\frac{1}{r^{2\triangle-2}}-\frac{2M}{r}, ~~~~\mathrm{if}~1<2\triangle\leq 3,\nonumber\\
&&g(r)\sim\frac{r^2}{\ell^2}-\frac{2M}{r}, ~~~~\mathrm{if}~3<2\triangle.
\eea
After using the scaling symmetries there are only two parameters at the horizon which can be used as initial values: $\eta_+$ and $E_+$. At the boundary, we have five parameters which give the properties of the dual field theory: $\mu$, $\rho$, $\psi^{(1)}$, $\psi^{(2)}$ and $M$. Thus by integrating out from horizon to infinity, we have a map:
\be\label{map} (\eta_+,E_+)\mapsto (\mu,\rho,\psi^{(1)},\psi^{(2)}, M).\ee

\section{Numerical Results for the Condensates}

In this section we give the numerical results for the condensates of the dual superconductor in a canonical ensemble, {\it i.e.} the charge density of the system is fixed to a value $\rho$. For the value of $m$ we choose, there can be two boundary conditions for $\eta(r)$: $\psi^{(1)}=0$ with $\psi^{(2)}$ giving the condensate ${\mathcal{O}}_2$ or $\psi^{(2)}=0$ with $\psi^{(1)}$ giving the value of the condensate ${\mathcal{O}}_1$. With the constraints from the boundary condition, the map (\ref{map}) reduces to a one parameter family of solutions for each choice of $\alpha$ and $q$. We can think of this parameter as being the temperature of the theory at a fixed charge density.

We find that for the values of $\alpha$ and $q$ we considered, there is always a critical temperature $T_c$ below which charged hairy black hole solutions can be found and above this critical temperature only AdS RN black hole solutions exist at a fixed nonzero charge density.

The free energy in the canonical system can be calculated from the
Euclidean gravity action through a Legendre transformation. The free energy of the hairy black hole is calculated to be \cite{Hartnoll:2008kx,{Cadoni:2009xm}}\be F_{\mathrm{Hairy}}=V(-2M+\mu\rho),\ee
and the free energy of the AdS RN black hole is\be F_{\mathrm{RN}}=\frac{V}
{r_+}(-r_+^4+\frac{3\rho^2}{4}),\ee where $V$ is the volume of
the $(x,y)$ plane.

\begin{figure}[h!]
\begin{center}
\begin{tabular}{cc}
\includegraphics[width=0.5\textwidth]{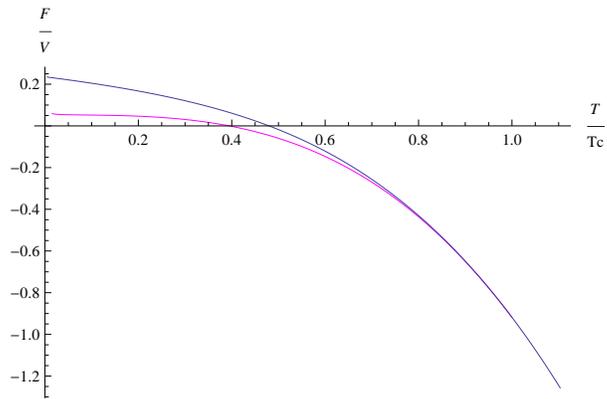}
\end{tabular}
\caption{\small  Free energies for the hairy black hole (the purple line)
and the AdS RN black hole (the blue line) at a fixe charge density
for $\alpha=5,q=1 $ and the operator ${\mathcal{O}}_1$. }
\label{fe}
\end{center}
\end{figure}

At temperatures lower than the critical temperature $T_c$, hairy black holes have a smaller free energy than the AdS RN black holes. As an
illustration we plot the picture of free energies for hairy black holes and AdS RN black holes for $\alpha=5$ and $q=1$ and the operator ${\mathcal{O}}_1$ in Figure \ref{fe}. Thus below this critical temperature, the system is in a superconducting phase while above this critical temperature,
the system is in a normal phase.

%%%%%%%%%%%%%%%%%%%%%%%%%%%%%%%%%%%%%%
%%%%%%%%%%%%%%%%%%%%%%%%%%%%%%%%%%%%%%
The origin of this instability of AdS RN black holes at low temperatures is the same as in the minimal model \cite{Hartnoll:2008kx}, which can be attributed to the fact that the effective mass of the scalar in the zero temperature limit of the AdS RN black holes violates the Breitenlohner-Freedman (BF) bound near the horizon.

For the AdS RN black hole (\ref{rn}), the temperature is $T=(12-\rho^2)/16\pi$, so the extremal limit is at $\rho=2\sqrt{3}$ and the near horizon geometry of the AdS RN black hole is AdS$_2 \times$ R$^2$, {\it i.e.}
\be\label{ads2} ds^2=-6(r-1)^2dt^2+\frac{dr^2}{6(r-1)^2}+dx^2+dy^2, ~~\phi=2\sqrt{3}(r-1).\ee

Plug (\ref{ads2}) into (\ref{eta}) and we can recover a wave equation in AdS$_2$ in the $\eta\ll 1$ limit,
\be \eta_{,\tilde{r}\tilde{r}}+\frac{2}{\tilde{r}}\eta_{,\tilde{r}}-
\frac{m_{\mathrm{eff}}^2}{\tilde{r}^2}\eta=0,\ee
where we introduced a new effective mass
\be m_{\mathrm{eff}}^2=\frac{m^2-2q^2-6\alpha^2}{6}, \ee
and a new variable $\tilde{r}=r-1$.

 An instability would arise when the mass of $\eta$ violates the BF bound in the near horizon region, {\it i.e.} the AdS$_2$ spacetime, while satisfies the BF bound for four dimensional AdS$_4$ spacetime:
\be m^2-2q^2-6\alpha^2<-\frac{3}{2},~~~~m^2>-\frac{9}{4}.\ee
This kind of instability is very useful to the realization of holographic phase transition, and a recent application can be found in \cite{Iqbal:2010eh}.

\begin{figure}[h!]
\begin{center}
\begin{tabular}{cc}
\includegraphics[width=1\textwidth, height=0.7\textwidth]{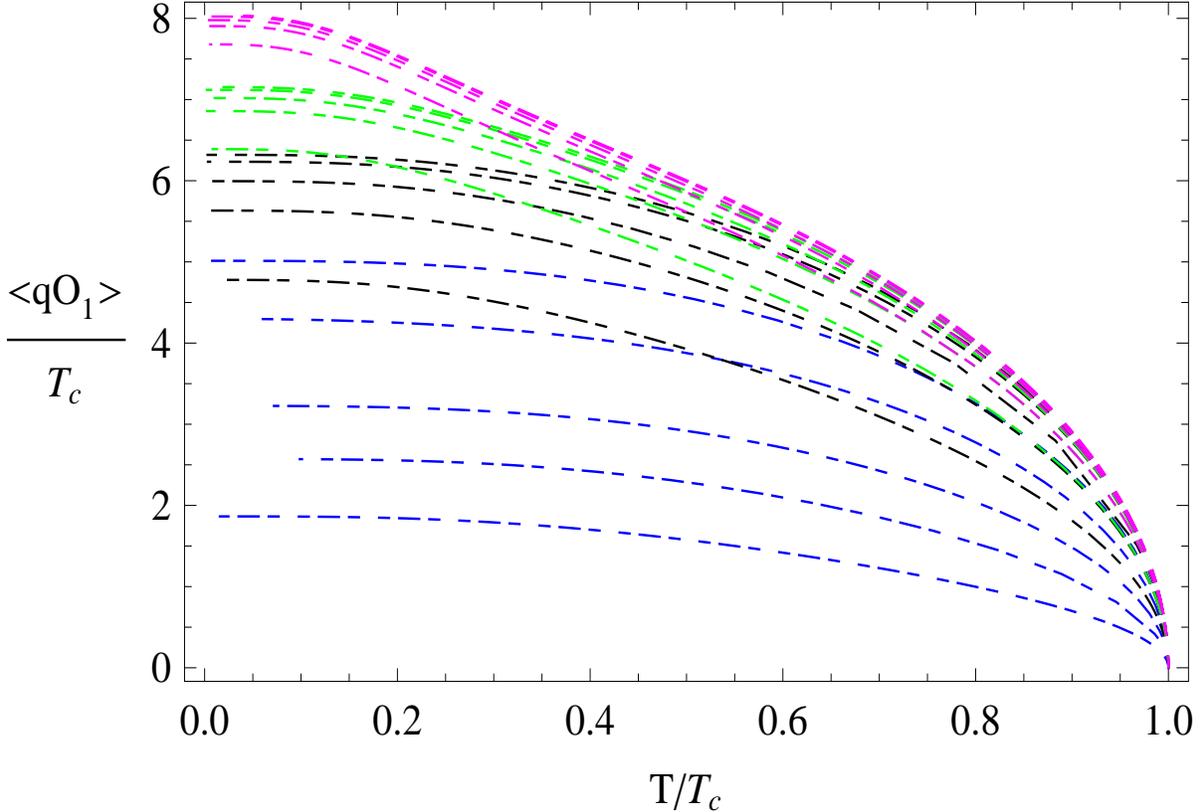}
\end{tabular}
\caption{\small  Values of the condensate ${\mathcal{O}}_1$ as a
function of the temperature for all the combinations of
$\alpha=0,1,2,3,5$, and
$q=1 ~(\mathrm{blue}),3 ~(\mathrm{black}),5 ~(\mathrm{green}),8 ~ (\mathrm{purple})$. For each color, the lines from top to down correspond to $\alpha=0,1,2,3,5$ with the same value of $q$ dictated by the
color respectively. }\label{o1}
\end{center}
\end{figure}

\begin{figure}[h!]
\begin{center}
\begin{tabular}{cc}
\includegraphics[width=1\textwidth, height=0.7\textwidth]{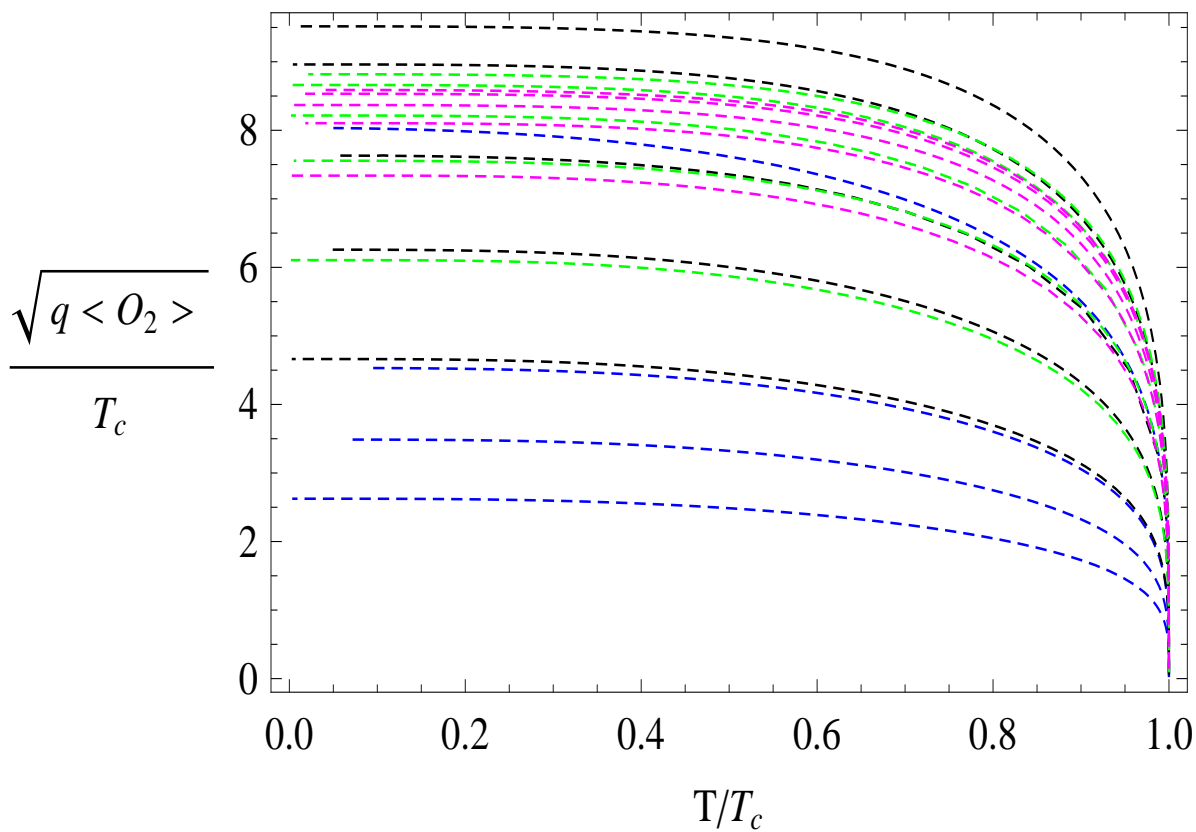}
\end{tabular}
\caption{\small  Values of the condensate ${\mathcal{O}}_2$
as a function of the temperature for all the combinations of
 $\alpha=0,1,2,3,5$, and $q=1~(\mathrm{blue}),3~(\mathrm{black}),5~(\mathrm{green}),8~(\mathrm{purple})$
 except $\alpha=0,~q=1$. For each color, the lines from top to down correspond to $\alpha=0,1,3,5,8$ with the same $q$ dictated by the color respectively. The case $\alpha=0,~q=1$ is not in the figure as it has a much
larger $\sqrt{\langle{\mathcal{O}}_2\rangle}/T_c$ (about 21) than other cases. }\label{o2}
\end{center}
\end{figure}

The values of the condensates in the superconducting phase
as functions of the temperature for various values
of $\alpha$ and $q$ are plotted
in Figure \ref{o1} for ${\mathcal{O}}_1$ and
in Figure \ref{o2} for ${\mathcal{O}}_2$ respectively.
As noticed in \cite{Cadoni:2009xm,{Horowitz:2010gk}},
for any given values of $\alpha$ and $q$ there are usually
several different hairy black hole solutions
with the same correct asymptotic behavior and we always
choose the only one solution with a monotonic scalar profile.

%%%%%%%%%%%%%%%%%%%%%%%%%%%%%%%%%%%%%%
%%%%%%%%%%%%%%%%%%%%%%%%%%%%%%%%%%%%%%

%%%%%%%%%%%%%%%%%%%%%%%%%%%%%%%%%%%%%%
%%%%%%%%%%%%%%%%%%%%%%%%%%%%%%%%%%%%%%

In Figure \ref{o1}, the values of the condensate
${\mathcal{O}}_1$ as a function of the temperature for
$\alpha=0,1,2,3,5$, and $q=1,3,5,8$ are plotted. It can
be easily seen that for any fixed value of $q$, the value
of ${\mathcal{O}}_1$ decreases as the value of $\alpha$ increases.
In Figure \ref{o2}, the values of ${\mathcal{O}}_2$ as a
function of the temperature for $\alpha=0,1,2,3,5$, and
$q=1,3,5,8$ are plotted. Note that as the maximum value
of ${\mathcal{O}}_2$ for $\alpha=0$ and $q=1$ is much
larger than for the other values of $\alpha$ and $q$, it is not
plotted out in Figure \ref{o2}. In Figure \ref{o2}, it can also
be discovered that as $\alpha$ increases, the value of
${\mathcal{O}}_2$ decreases for fixed values of $q$.

It can be checked that $\frac{\partial F_{\mathrm{Hairy}}}{V\partial T}\bigg{|}_{T=T_c}=\frac{\partial F_{\mathrm{RN}}}{V\partial T}\bigg{|}_{T=T_c}$ while $\frac{\partial^2 F_{\mathrm{Hairy}}}{V\partial T^2}\bigg{|}_{T=T_c}\neq\frac{\partial^2 F_{\mathrm{RN}}}{V\partial T^2}\bigg{|}_{T=T_c}$, so the phase transition at $T_c$ is a
second order phase transition. We can also see this from the behavior
of the condensates near the critical temperature $T_c$. Near $T_c$,
both the condensates ${\mathcal{O}}_1$ and ${\mathcal{O}}_2$
behave like $q{\mathcal{O}}_i\approx a_iT_c^i(1-T/T_c)^{1/2},$ for $i=1,2$ and
$a_1$ and $a_2$ are two constants which differ while $\alpha$ and $q$
change. In Table \ref{a1a2} we list the values of $a_1$ and $a_2$
corresponding to different values of $\alpha$ and $q$. The behavior
$q{\mathcal{O}}_i\approx a_iT_c^i(1-T/T_c)^{1/2},$ for $i=1,2$ is consistent with the prediction from the mean field theory for second order phase transitions \cite{Hartnoll:2008vx}.

\begin{minipage}{\textwidth}
\begin{minipage}[t]{0.5\textwidth}
\begin{tabular}{|c||c|c|c|c|c|c|c|}
\hline
$q\backslash \alpha$ & 0 & 1 & 2 & 3 & 5 \\
\hline
\hline
1 & 8.0 & 6.7 & 4.7 & 3.4 & 2.2 \\
\hline
3 & 9.0 & 8.7 & 8.4 & 7.5 & 5.7 \\
\hline
5 & 9.1 & 8.9 & 8.7 & 8.6 & 7.6 \\
\hline
8 & 9.3 & 9.3 & 9.3 & 9.0 & 8.6
 \\ \hline
\end{tabular}
\end{minipage}
\begin{minipage}[t]{0.5\textwidth}
\begin{tabular}{|c||c|c|c|c|c|}
\hline
 $q\backslash \alpha$ & 0&1&2&3&5\\
\hline
\hline
1 & 670 & 100 &30 & 17 & 9.3 \\
\hline
3 & 182 & 153 & 97& 60 & 31 \\
\hline
5 & 155 & 146 &126&  98 & 57 \\
\hline
8 & 148 & 145 & 137&122 & 92\\
\hline
\end{tabular}
\end{minipage}\makeatletter\def\@captype{table}\makeatother
\caption{\small  Left: The coefficient $a_1$ for various combinations
of $\alpha$ and $q$. Right: The coefficient $a_2$ for various
combinations of $\alpha$ and $q$.}
\label{a1a2}
\end{minipage}

\begin{figure}[h!]
\begin{center}
\begin{tabular}{cc}
\includegraphics[width=0.5\textwidth]{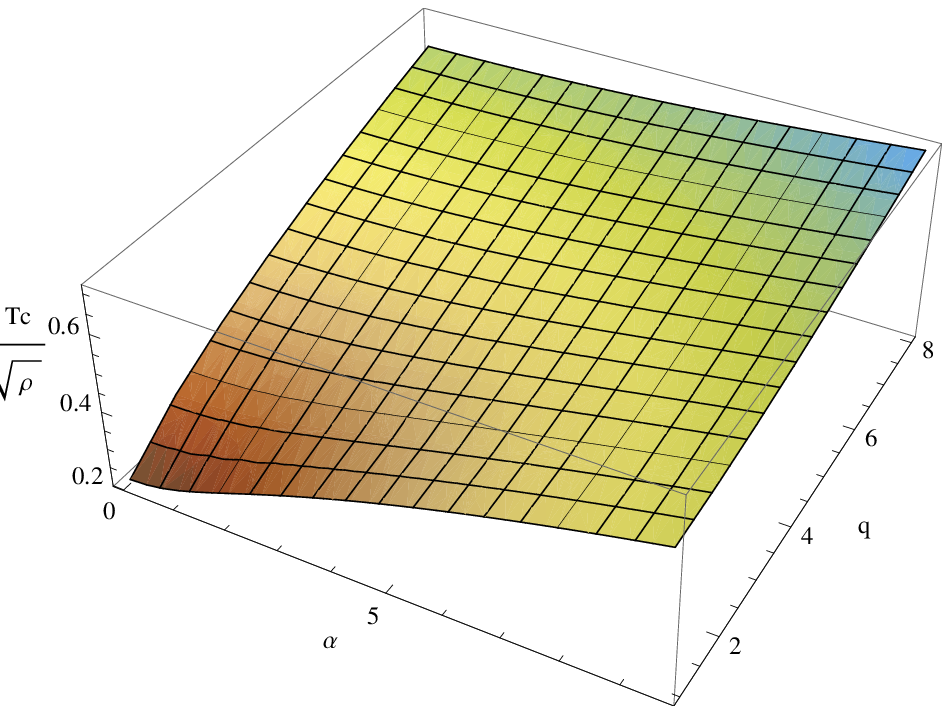}
\includegraphics[width=0.5\textwidth]{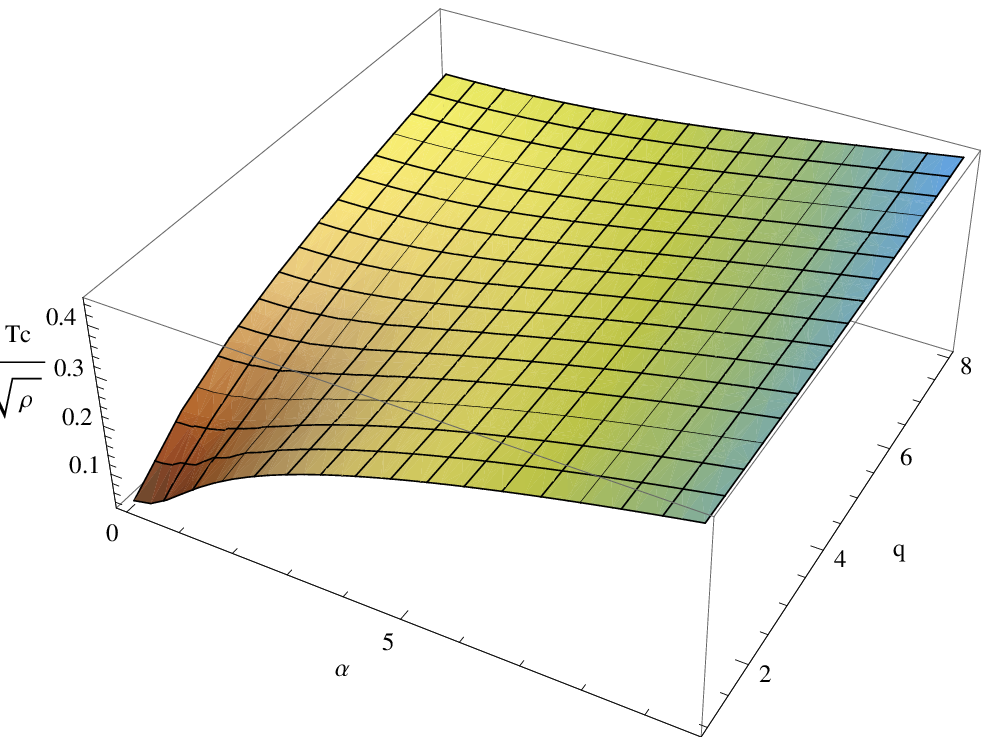}
\end{tabular}
\caption{\small  Left: The values of $T_c/\sqrt{\rho}$ as
a function of $\alpha$ and $q$ for the operator ${\mathcal{O}}_1$;
Right: The values of $T_c/\sqrt{\rho}$ as a function of $\alpha$
and $q$ for the operator ${\mathcal{O}}_2$.}\label{falpha1}
\end{center}
\end{figure}

The critical temperature $T_c$ is proportional to $\sqrt{\rho}$
and it also depends on the values of $\alpha$ and $q$. $T_c/\sqrt{\rho}$
increases as $\alpha$ or $q$ increases. In Figure \ref{falpha1}, the values of $T_c/\sqrt{\rho}$ as functions of $\alpha$ and $q$ are
plotted for the operators ${\mathcal{O}}_1$ and ${\mathcal{O}}_2$
respectively.

%%%%%%%%%%%%%%%%%%%%%%%%%%%%%%%%%%%%%%
%%%%%%%%%%%%%%%%%%%%%%%%%%%%%%%%%%%%%%

\section{Electric Conductivity}
The electric conductivity of the dual superconductor in the
superconducting phase can be calculated from linear perturbations
of $g_{tx}$ and $A_x$ around the hairy black hole in the gravity
side. We consider perturbations with zero momentum:
 $A_x=a_x(r)e^{-i\omega t}$ and
$g_{tx}=f(r)e^{-i\omega t}$, and these perturbations can get
decoupled from other perturbations. The equations of motion for
the two perturbations are
\bea
\label{flucax}&&a_x''+\bigg(\frac{g'}{g}-\frac{\chi'}{2}+\frac{\partial_\eta
G}{G}\eta'\bigg)a_x'
+\bigg(\frac{\omega^2}{g^2}e^{\chi}-\frac{J}{gG}\bigg)a_x
+\frac{\phi'}{g}e^{\chi}\bigg(f'-\frac{2}{r}f\bigg)=0,
\\\label{flucgxt}&& f'-\frac{2}{r}f+G\phi'a_x=0.\eea

Plug (\ref{flucgxt}) into (\ref{flucax}), and we can obtain
a single equation of motion for $a_x(r)$:
\be\label{eomax}
a_x''+\bigg(\frac{g'}{g}-\frac{\chi'}{2}+\frac{\partial_\eta
G}{G}\eta'\bigg)a_x'
+\bigg(\big(\frac{\omega^2}{g^2}-\frac{G\phi'^{2}}{g}\big)e^{\chi}-\frac{J}{gG}\bigg)a_x=0.
\ee

The asymptotic behavior of the Maxwell field near the
boundary is \be a_x\sim
a_x^{(0)}+\frac{a_x^{(1)}}{r},\ee
and the conductivity of the dual superconductor
can be calculated from the formula \cite{Hartnoll:2008kx}
\be\label{conduc1} \sigma(\omega)=-\frac{i}{\omega}\frac{a_x^{(1)}}{a_x^{(0)}}.\ee

Thus in order to get the electric conductivity,
we still have to use numerical calculations to get the values of
$a_x^{(0)}$ and $a_x^{(1)}$ on hairy black hole backgrounds
with different temperatures for various combinations of $\alpha$
and $q$. Before  performing the numerical calculations,
we need to get the behavior of $a_x$ near the horizon. From (\ref{eomax}) it can be easily seen that $a_x$ should
vanish as power law of $g(r)$ near the horizon, {\it i.e}.
\be a_x\propto g^{-i\omega \sqrt{\frac{e^{\chi}}{g'^2}}\big|_{r=r_+}}.\ee

%%%%%%%%%%%%%%%%%%%%%%%%%%%%%%%%%%%%%%
\begin{figure}[h!]
\begin{center}
\begin{tabular}{ccc}
  \includegraphics[width=0.3\textwidth]{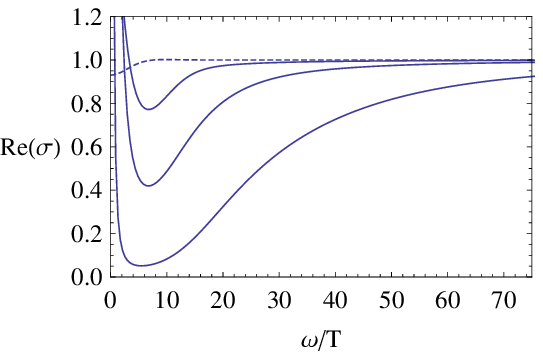}
 \includegraphics[width=0.3\textwidth]{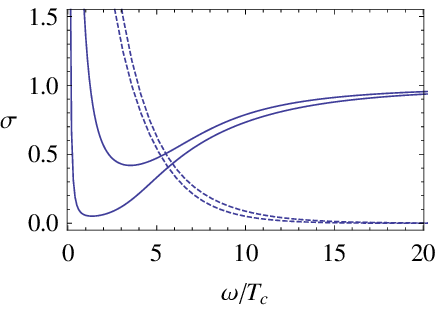}
   \includegraphics[width=0.3\textwidth]{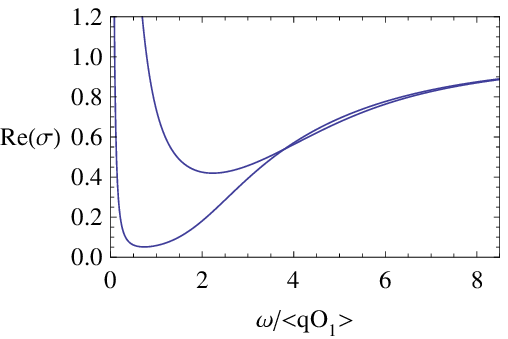}
  \end{tabular}
  \begin{tabular}{ccc}
   \includegraphics[width=0.3\textwidth]{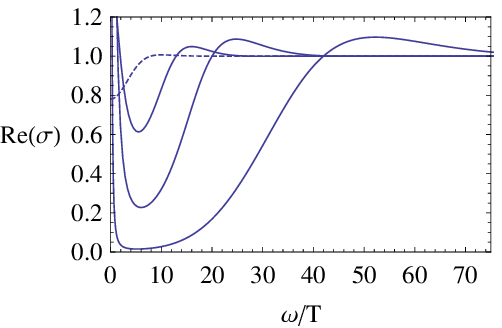}
  \includegraphics[width=0.3\textwidth]{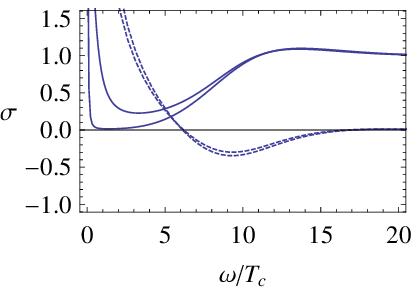}
  \includegraphics[width=0.3\textwidth]{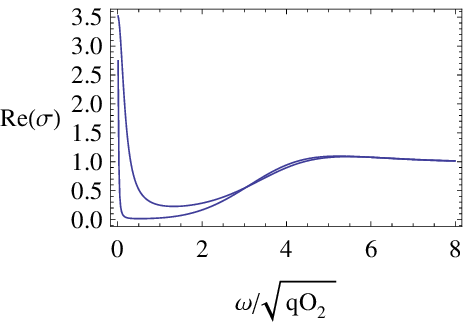}
  \end{tabular}
  \caption{\small  $\alpha=5,q=1$. The three pictures on top are electric
  conductivities for the condensate ${\mathcal{O}}_1$. From left to right, the real part of the electric conductivities are functions
  of $w/T$, $w/T_c$ and $w/q{\langle\mathcal{O}}_1\rangle$ respectively. In the middle, the imaginary part of the conductivity is also plotted
  using dashed lines. In the first picture, the lines correspond to
   to $T/T_c=1,0.801,0.499,0.201$ from top to bottom while in the second and the third pictures, the lines correspond to $T/T_c=0.499,0.201$ from top to bottom.
   The three pictures on the bottom are
   for the condensate ${\mathcal{O}}_2$. In the first picture, the lines correspond to $T/T_c=1, 0.801, 0.504, 0.200$ from top to bottom while in the second and the third pictures, the lines correspond to $T/T_c=0.504, 0.200$ from top to bottom. There is a delta
   function in ${\mathrm{Re}}(\sigma)$ at $w=0$, which is not plotted out. }\label{s51}
\end{center}
\end{figure}
%%%%%%%%%%%%%%%%%%%%%%%%%%%%%%%%%%%%%%

%%%%%%%%%%%%%%%%%%%%%%%%%%%%%%%%%%%%%%%
\begin{figure}[h!]
\begin{center}
\begin{tabular}{cc}
  \includegraphics[width=0.3\textwidth]{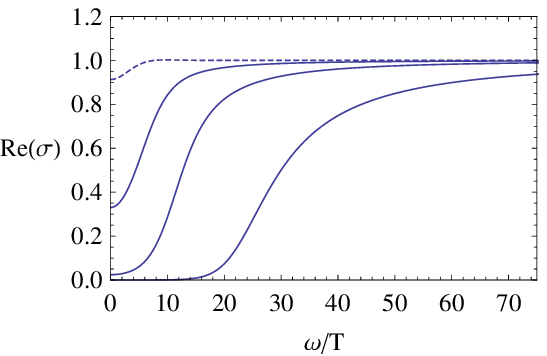}
  \includegraphics[width=0.3\textwidth,bb=0 0 153 107]{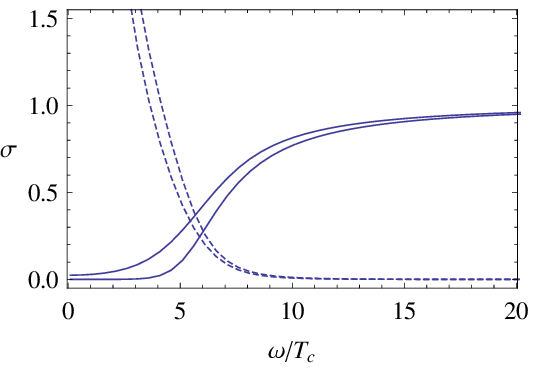}
   \includegraphics[width=0.3\textwidth]{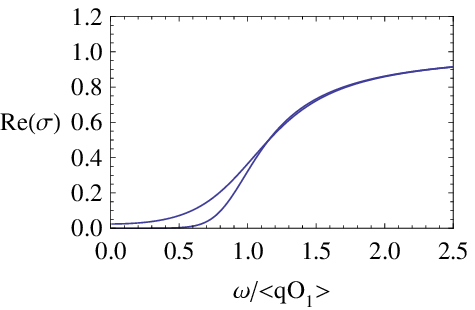}
    \end{tabular}
  \begin{tabular}{cc}
  \includegraphics[width=0.3\textwidth]{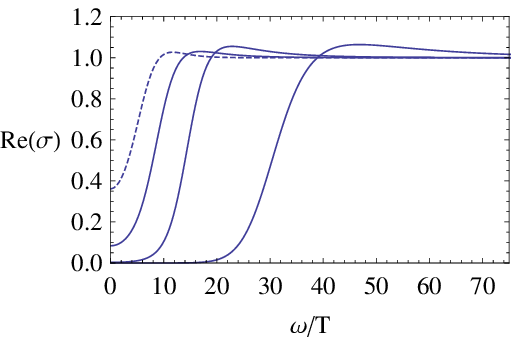}
  \includegraphics[width=0.3\textwidth]{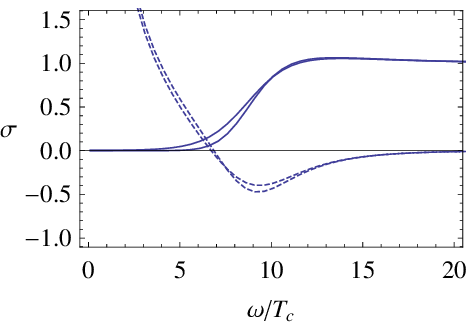}
  \includegraphics[width=0.3\textwidth]{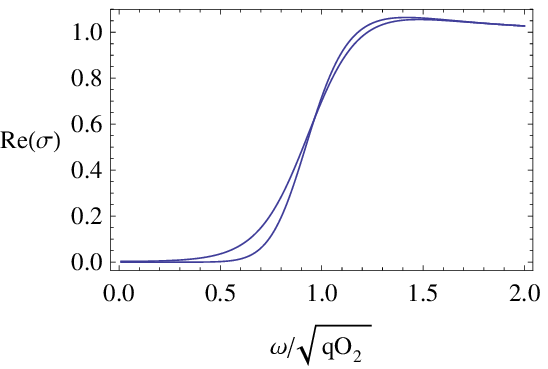}
  \end{tabular}
  \caption{\small  $\alpha=0,q=3$. The three pictures on top are electric
  conductivities for the condensate ${\mathcal{O}}_1$. In the first one, the lines correspond
   to $T/T_c=1,0.792,0.490,0.204$ from top to the bottom.
   The three pictures on the bottom are for the condensate ${\mathcal{O}}_2$. The lines in the first one correspond to $T/T_c=1, 0.804, 0.508, 0.205$ from top to bottom.
  In the four pictures on the right, the lines correspond to $T/T_c=0.5,0.2$ from top to bottom.}\label{s031}
\end{center}
\end{figure}
%%%%%%%%%%%%%%%%%%%%%%%%%%%%%%%%%%%%%%%

%%%%%%%%%%%%%%%%%%%%%%%%%%%%%%%%%%%%%%
\begin{figure}[h!]
\begin{center}
\begin{tabular}{cc}
  \includegraphics[width=0.3\textwidth]{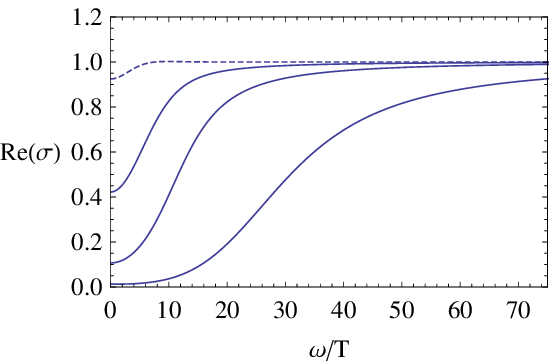}
  \includegraphics[width=0.3\textwidth,bb=0 0 121 87]{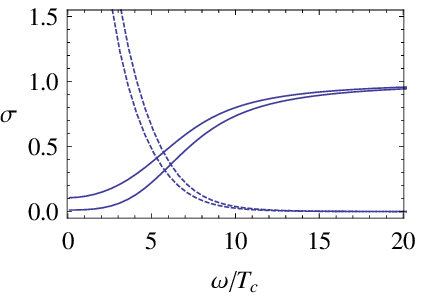}
    \includegraphics[width=0.3\textwidth]{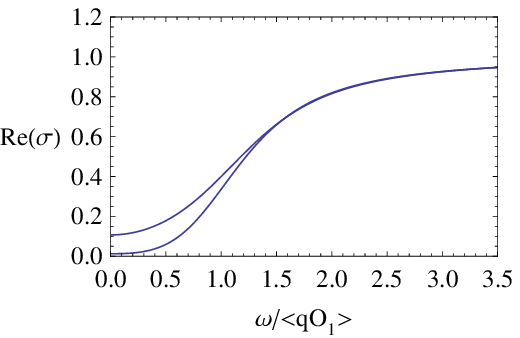}
    \end{tabular}
  \begin{tabular}{cc}
   \includegraphics[width=0.3\textwidth]{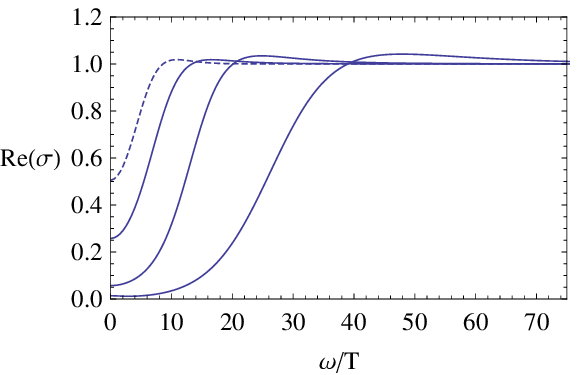}
  \includegraphics[width=0.3\textwidth]{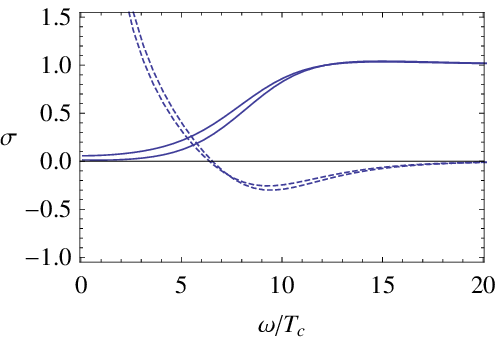}
  \includegraphics[width=0.3\textwidth]{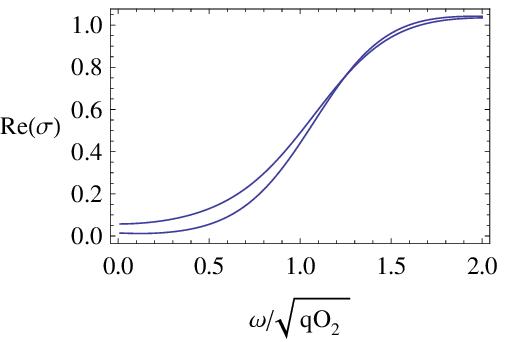}
  \end{tabular}
  \caption{\small  $\alpha=2,q=3$.
  The three pictures on top are electric
  conductivities for the condensate ${\mathcal{O}}_1$. The lines in the first picture correspond
   to $T/T_c=1,0.78,0.51,0.20$ from top to the bottom.
   The three pictures on the bottom are for the condensate ${\mathcal{O}}_2$. The lines in the first one correspond to $T/T_c=1, 0.84, 0.52, 0.21$ from top to bottom. In the four pictures on the right, the lines correspond to $T/T_c=0.5,0.2$ from top to bottom.}\label{s23}
\end{center}
\end{figure}
%%%%%%%%%%%%%%%%%%%%%%%%%%%%%%%%%%%%%%

%%%%%%%%%%%%%%%%%%%%%%%%%%%%%%%%%%%%%%
\begin{figure}[h!]
\begin{center}
\begin{tabular}{cc}
  \includegraphics[width=0.3\textwidth]{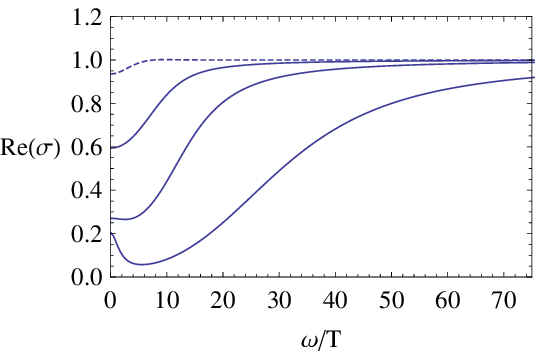}
    \includegraphics[width=0.3\textwidth]{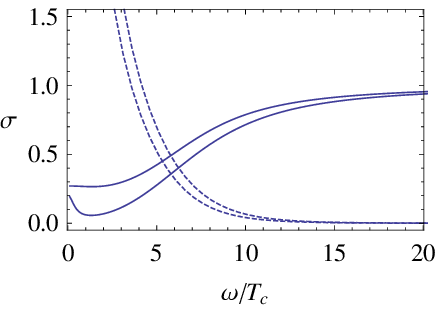}  \includegraphics[width=0.3\textwidth]{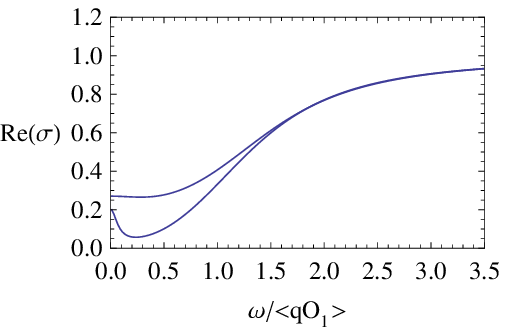}
      \end{tabular}
  \begin{tabular}{cc}
  \includegraphics[width=0.3\textwidth]{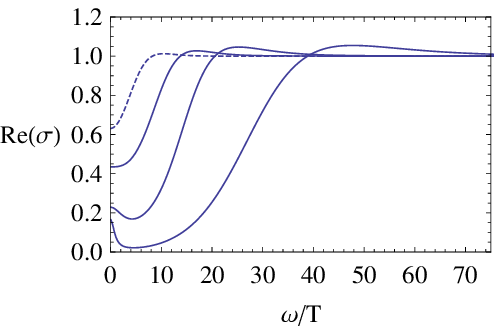}
  \includegraphics[width=0.3\textwidth]{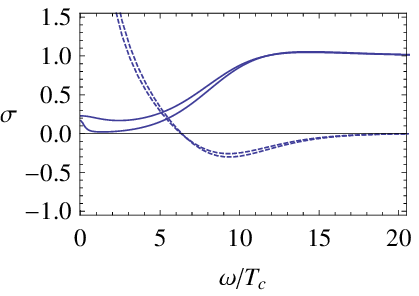}
  \includegraphics[width=0.3\textwidth]{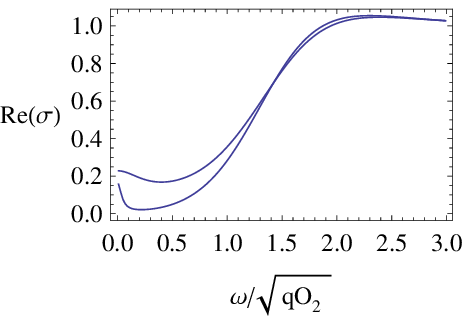}
  \end{tabular}
  \caption{\small  $\alpha=3,q=3$. The three pictures on top are electric
  conductivities for the condensate ${\mathcal{O}}_1$. The lines in the first picture correspond
   to $T/T_c=1,0.8,0.5,0.2$ from top to the bottom.
   The three pictures on the bottom are for the condensate ${\mathcal{O}}_2$. The lines in the first picture correspond to $T/T_c=1, 0.8, 0.5, 0.2$ from top to bottom. In the four pictures on the right, the lines correspond to $T/T_c=0.5,0.2$ from top to bottom.}\label{s33}
\end{center}
\end{figure}
%%%%%%%%%%%%%%%%%%%%%%%%%%%%%%%%%%%%%%

%%%%%%%%%%%%%%%%%%%%%%%%%%%%%%%%%%%%%%
\begin{figure}[h!]
\begin{center}
\begin{tabular}{ccc}
  \includegraphics[width=0.3\textwidth]{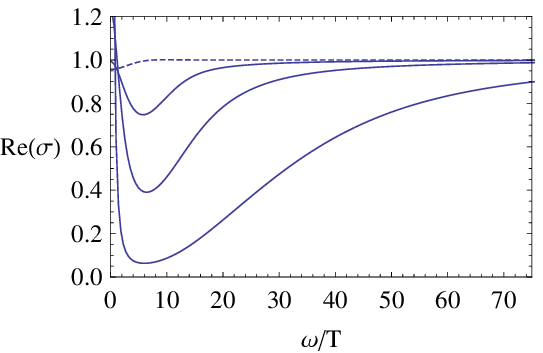}
      \includegraphics[width=0.3\textwidth]{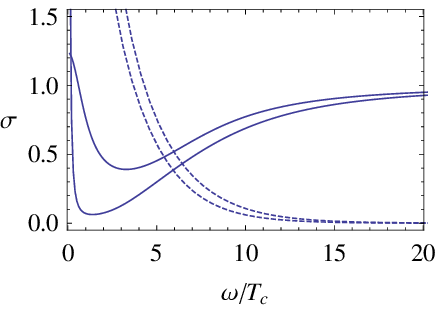}
  \includegraphics[width=0.3\textwidth]{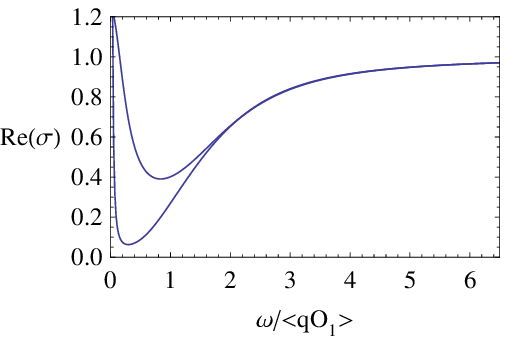}
  \end{tabular}
  \begin{tabular}{ccc}
      \includegraphics[width=0.3\textwidth]{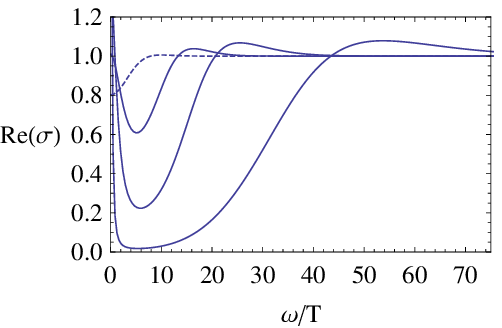}
  \includegraphics[width=0.3\textwidth]{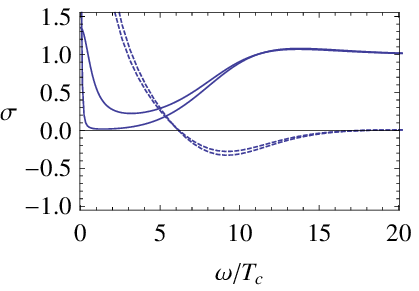}
  \includegraphics[width=0.3\textwidth]{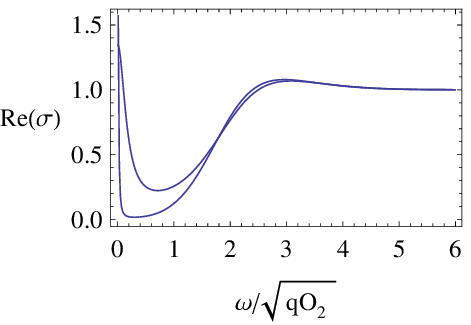}
  \end{tabular}
  \caption{\small  $\alpha=5,q=3$. The top three pictures are for the condensate ${\mathcal{O}}_1$ and the bottom for ${\mathcal{O}}_2$. The lines
  in the two pictures on the left correspond to $T/T_c=1,0.8,0.5,0.2$ from top to bottom. In the four pictures on the right, the lines correspond to $T/T_c=0.5,0.2$ from top to bottom.}\label{s53}
\end{center}
\end{figure}
%%%%%%%%%%%%%%%%%%%%%%%%%%%%%%%%%%%%%%

%%%%%%%%%%%%%%%%%%%%%%%%%%%%%%%%%%%%%%
\begin{figure}[h!]
\begin{center}
\begin{tabular}{ccc}
  \includegraphics[width=0.3\textwidth]{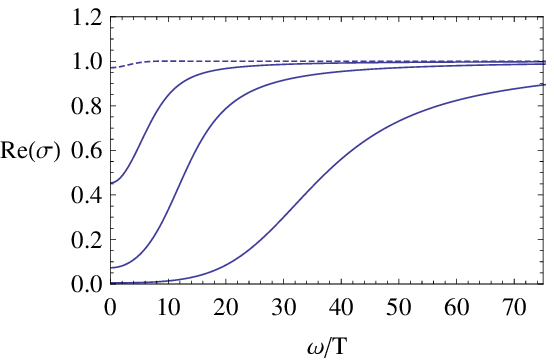}
     \includegraphics[width=0.3\textwidth,bb=0 0 139 98]{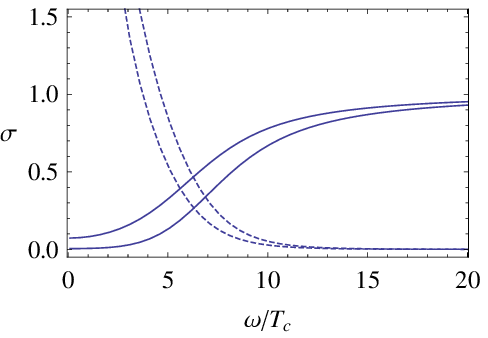}
  \includegraphics[width=0.3\textwidth]{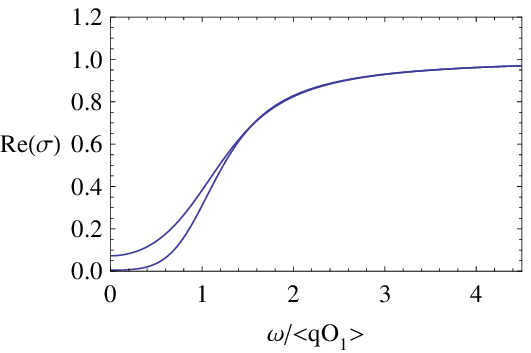}
  \end{tabular}
  \begin{tabular}{ccc}
      \includegraphics[width=0.3\textwidth]{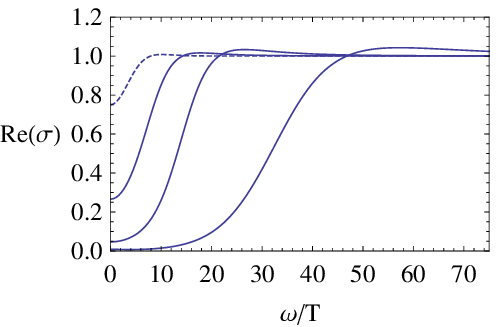}
  \includegraphics[width=0.3\textwidth]{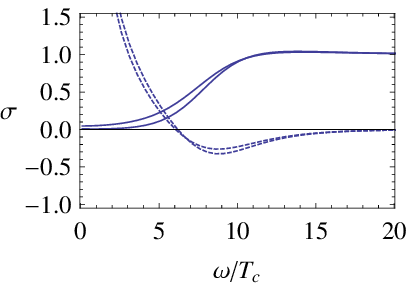}
  \includegraphics[width=0.3\textwidth]{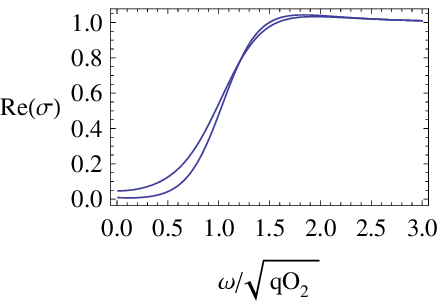}
  \end{tabular}
  \caption{\small   $\alpha=3,q=5$. The top three pictures are for the condensate ${\mathcal{O}}_1$ and the bottom are for ${\mathcal{O}}_2$. The lines
  in the two pictures on the left correspond to $T/T_c=1,0.8,0.5,0.2$ from top to bottom. In the four pictures on the right, the lines correspond to $T/T_c=0.5,0.2$ from top to bottom.}\label{s35}
\end{center}
\end{figure}
%%%%%%%%%%%%%%%%%%%%%%%%%%%%%%%%%%%%%%
The values of electric conductivities as functions of
$\omega/T$, $\omega/T_c$ and $\omega/(q{\langle\mathcal{O}}_i\rangle)^{1/i},~i=1,2$ are plotted in Figures \ref{s51}, \ref{s031}, \ref{s23}, \ref{s33}, \ref{s53}, \ref{s35} for $(\alpha,q)=(5,1),(0,3),(2,3),(3,3),(5,3),(3,5)$ respectively. As there is a pole at $\omega=0$ in the imaginary part
of the electric conductivity, a delta function arises in the real part
of the electric conductivity at $\omega=0$ using the Kramers-Kronig relations.

From these figures, we can see that at fixed values
of $q$, the curves of the electric conductivities behave similarly to the minimal model (Figure \ref{s031}) in \cite{Hartnoll:2008kx} when $\alpha$ is small, and as we increase the value of $\alpha$ some novel behavior of the electric conductivities arises. As an illustration, we can compare Figure \ref{s53} with Figure \ref{s031} and find that at the small $\omega$ region, the real parts of the conductivities of the two models behave quite differently. In Figure \ref{s53} with $\alpha=5$ the real part of the conductivity exhibits a ``Drude Peak" in the region $\omega\to 0$ and has a dip at a finite and small value of $\omega$. Among these figures, Figure \ref{s51} for $\alpha=5$ and $q=1$ has the largest deviation from the minimal model and it behaves similarly to Figure \ref{s53} qualitatively while it has larger maximum values in the limit $\omega\to 0$, which are $~0.93, ~1.82, ~7.2, ~23$ for ${
\mathcal{O}}_1$ and $0.78, ~1.44, ~3.50, ~3.0$ for ${
\mathcal{O}}_2$ in Figure \ref{s51}. This novel behavior has also been found in holographic strange metals by \cite{Hartnoll:2009ns} and in \cite{Cadoni:2009xm} for the case $q=0$.

\begin{figure}[h!]
\begin{center}
\begin{tabular}{cc}
\includegraphics[width=0.5\textwidth]{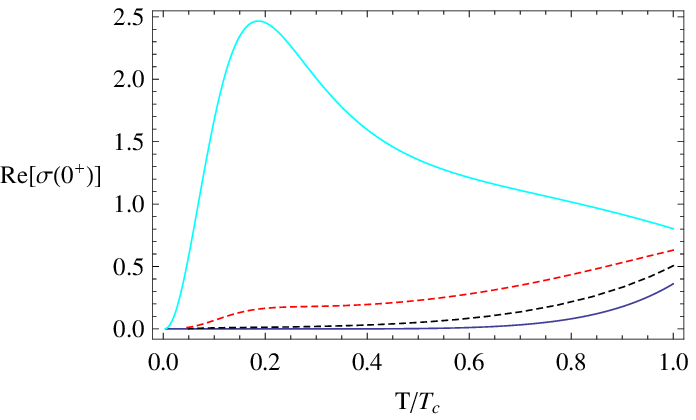}
\includegraphics[width=0.5\textwidth]{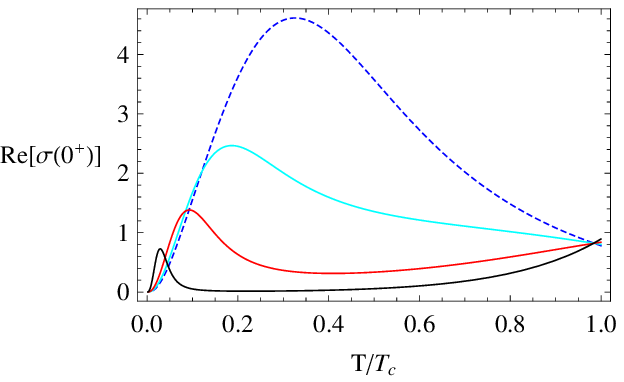}
\end{tabular}
\caption{\small  Left: The height of the peak at $\omega\to 0$ for $q=3$, $\alpha=$ 0 (blue), 2 (black), 3 (red), 5 (cyan) for the operator ${\mathcal{O}}_2$; Right:  The height of the peak at $\omega\to 0$ for $\alpha=5$, $q=$ 1 (blue), 3 (cyan), 5 (red), 8 (black) for the operator ${\mathcal{O}}_2$.}\label{dc}
\end{center}
\end{figure}

As there are peaks for Re$(\sigma)$ at $\omega\to 0$ for certain values of $\alpha$ and $q$, the height of the peaks depends on the temperature and $\alpha$, $q$. In figure \ref{dc} we show the dependence of the height on the temperature for some values of $\alpha$ and $q$. We can see that for $\alpha=5$, the height does not increase monotonically like in the minimal model and has a maximum at some finite temperature. This maximum value also decreases as $q$ increases. Thus we can see that all these novel properties
 at small frequencies become more apparent as $\alpha/q$ goes larger.

%%%%%%%%%%%%%%%%%%%%%%%%%%%%%%%%%%%%%%
%%%%%%%%%%%%%%%%%%%%%%%%%%%%%%%%%%%%%%

From Figure \ref{dc}, we can also find that as the temperature approaches zero, {\it i.e.} $T/T_c\to 0$, the height of the peak vanishes for all the values of $\alpha$ and $q$ considered here. Thus as $T/T_c\to 0$, the peak vanishes and a gap arises, so the behavior of Re$(\sigma)$ becomes again similar to the case of the minimal model. From the Figures \ref{s51}, \ref{s031}, \ref{s23}, \ref{s33}, \ref{s53}, \ref{s35} we can also see that the imaginary parts of the conductivities exhibit poles at $\omega=0$ and have minimums at $\omega/T_c\approx 8$ for the operator ${\mathcal{O}}_2$, so we can also expect here that the width of the gap $\omega_g/T_c \approx 8$  as $T/T_c\to 0$, which is also the same as in the minimal model \cite{Horowitz:2008bn}.

It is apparent that when we adjust the values of $\alpha$ and $q$ we can get the curves of electric conductivity with quite different shapes and it is expected that for certain values of $\alpha$ and $q$ we can get very close to the shapes of the curves observed in the experiments, for example, Fig. 14 in \cite{basov}.

The different small frequency behavior of electric conductivities can also be shown by rewriting the equation (\ref{ax}) into a one dimensional Schr\"{o}dinger equation and expressing the electric conductivity using the reflection coefficient. To do this, we introduce a new radial variable $u$ which is defined by \be
 du=\frac{e^{\chi/2}}{g}dr.\ee At large $r,$ $du=dr/r^2$, so
 $u=-1/r$. In this new coordinate system, the boundary is at $u=0$ and the horizon corresponds to $u=-\infty.$ We can define a new field $\Psi=\sqrt{G}a_x$, and (\ref{eomax}) becomes a one-dimensional
 Schr\"{o}dinger equation in the new coordinate system, \be\label{schequ1}
 \frac{d^2\Psi}{du^2}+\bigg[\omega^2-V(u)\bigg]\Psi=0,\ee with
 \be\label{potential}
 V(u)=g(G\phi'^2+\frac{J}{G}e^{-\chi})+\frac{1}{\sqrt{G}}
 \frac{d^2\sqrt{G}}{du^2}
 %+g^2e^{-\chi}\bigg(\frac{G''}{2G}-\frac{G'^2}{4G^2}
 %+\frac{G'}{2G}(\frac{g'}{g}-\frac{\chi'}{2})\bigg)
 ,\ee where the prime denotes the derivative with respect to $r$. The
 potential $V(u)$ has an extra contribution compared to the case of the minimal model.

To solve the equation (\ref{schequ1}) with ingoing wave boundary conditions at $u=-\infty$, we can first extend the definition of the potential to all $u$ by setting $V(u)=0$ for $u>0$ and then this equation can be solved through a one-dimensional scattering problem.
$\Psi(u)$ can be taken as the wave function and we consider an incoming wave from the right. Thus the transmitted wave is purely ingoing at the horizon, which satisfies our boundary condition. While at $u\geq 0,$ the wave function is
\be \Psi(u)=e^{-i\omega u}+{\mathcal{R}}e^{i\omega u},~~~ u\geq 0,\ee
where ${\mathcal{R}}$ is the reflection coefficient. Using the definition $\Psi=\sqrt{G}a_x$ and (\ref{conduc1}), we can have
\be\sigma(\omega)=\frac{1-{\mathcal{R}}}{1+{\mathcal{R}}}-\frac{i}{2\omega}\bigg(\frac{1}{G}
\frac{dG}{du}\bigg)\bigg|_{u=0}.
\ee
For the case $G=\mathrm{const}$, the formula above goes back to the original result obtained in \cite{Horowitz:2009ij}.
Note that the second term is purely imaginary and will not affect the real part of the conductivity. In fact, it vanishes after considering the asymptotical behavior of the scalar field $\eta$ and we can drop it out from the formula. Thus we can see that the real part of the conductivity is fully determined by the reflection coefficient ${\mathcal{R}}$, hence by the potential in the Schr\"{o}dinger equation.

The shape of $V(u)$ is crucial to the value of the reflection constant and we can analyze the property of $V(u)$ near the horizon and the boundary. Using the near boundary $u\to 0$ behavior of the fields we can obtain the near boundary behavior of the potential as
\be V(u)\sim \rho^2u^2+\bigg[\frac{\alpha^2}{2}\triangle(2\triangle-1)+q^2\bigg]
{\psi^{(\triangle)}}^2(-u)^{2(\triangle-1)}.\ee
Thus it vanishes for $\triangle>1$, and is a nonzero constant for $\triangle=1$ while diverges for $1/2<\triangle<1$. Near the horizon $u\to-\infty$, the first term in
(\ref{potential}) dominates and the potential vanishes as $V_he^{4\pi Tu}$. Thus in the following, we mainly focus on the behavior of $V(u)$ for the operator $\mathcal{O}_2$.

\begin{figure}[h!]
\begin{center}
\begin{tabular}{cc}
\includegraphics[width=0.5\textwidth, height=0.5\textwidth]{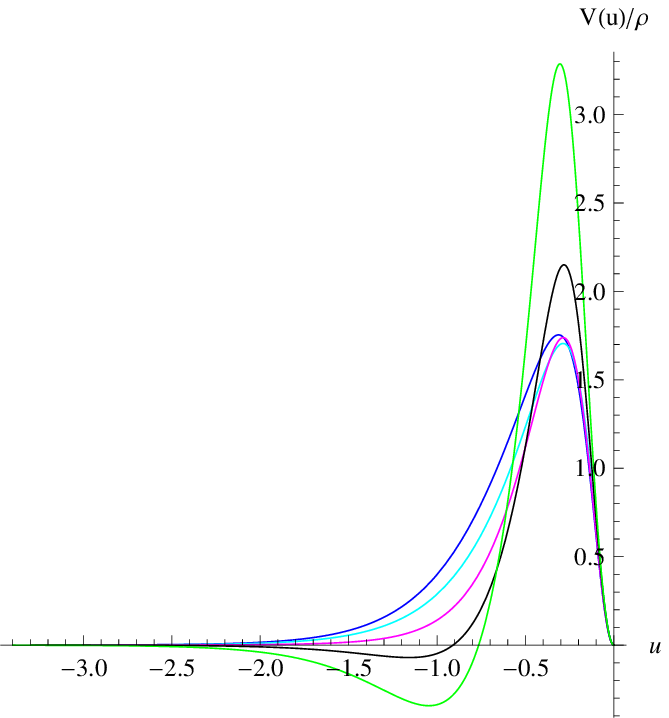}
\includegraphics[width=0.5\textwidth,height=0.5\textwidth]{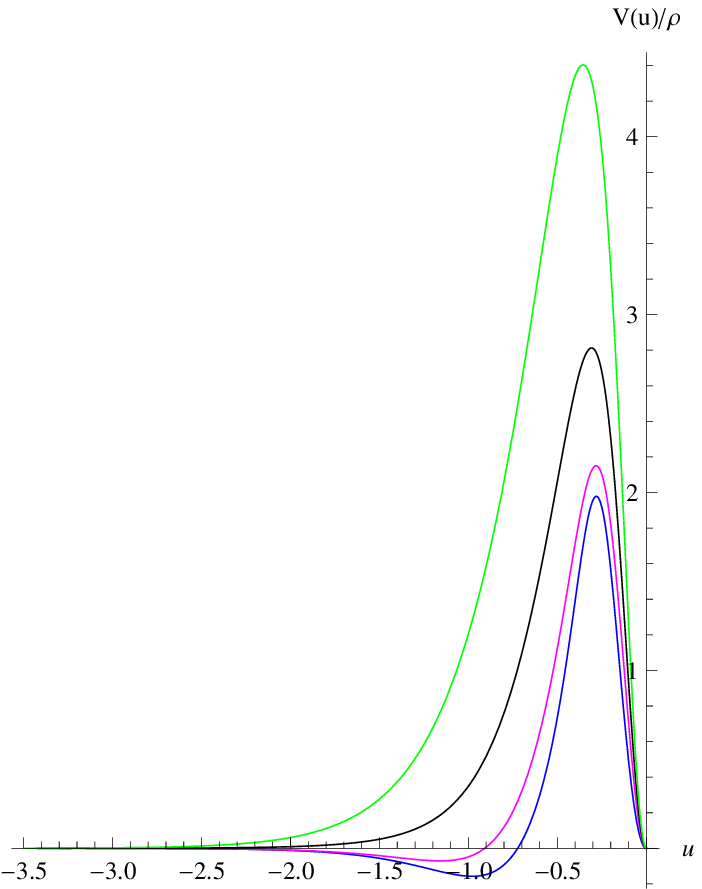}
\end{tabular}
\caption{\small V(u) for the operator ${\mathcal{O}}_2$. Left: $V(u)$ for q=3 and $\alpha=0$ (blue), $\alpha=1$ (cyan), $\alpha=2$ (purple), $\alpha=3$ (black), $\alpha=5$ (green) respectively, at $T/T_c=0.5$. Right: $V(u)$ for $\alpha=3$ and $q=1$ (blue), $q=3$ (purple), $q=5$ (black), $q=8$ (green) respectively, at $T/T_c=0.5$.}\label{vu}
\end{center}
\end{figure}

\begin{figure}[h!]
\begin{center}
\begin{tabular}{cc}
\includegraphics[width=0.4\textwidth, height=0.5\textwidth]{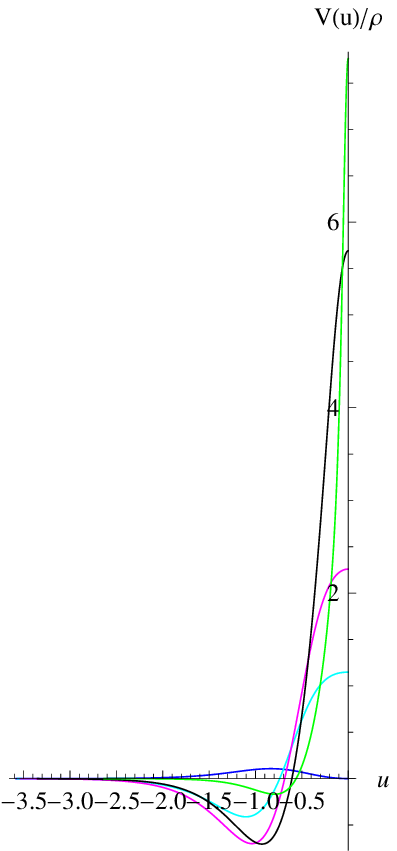}
\includegraphics[width=0.4\textwidth, height=0.5\textwidth,bb=0 0 164 202]{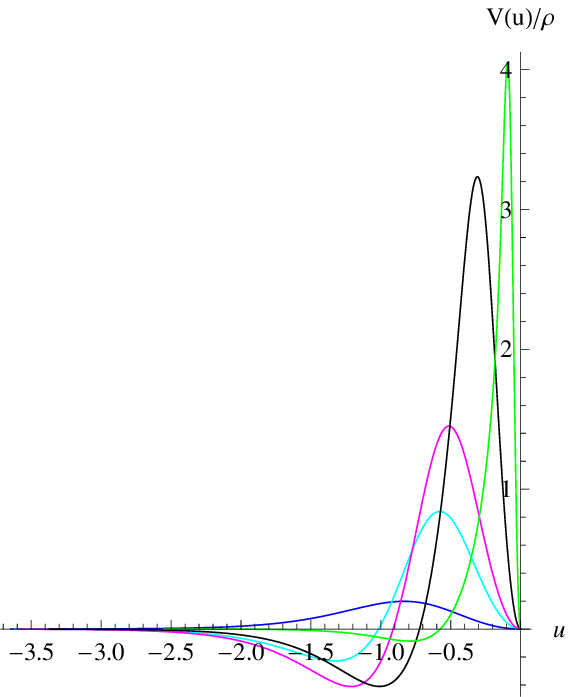}
\end{tabular}
\caption{\small   $V(u)$ for $\alpha=5,q=1$. Left: for the operator ${\mathcal{O}}_1$; Right: for the operator ${\mathcal{O}}_2$. Both of them are for $T/T_c=$ 1 (blue), 0.9 (cyan),  0.8 (purple), 0.5 (black), 0.2 (green). }\label{vu51}
\end{center}
\end{figure}

In Figures \ref{vu} and \ref{vu51} we show the shapes of the potential $V(u)$ for some values of $\alpha$ and $q$. In the minimal model, $V(u)$ is always positive along the $u$-axis and its peak becomes wider and higher as the temperature lowers. Thus in the minimal model the real part of the conductivity approaches $1$ at large frequencies and becomes very small at low frequencies. As the temperature lowers, the value of Re$(\sigma)$ at low frequencies also gets smaller. In our model, as $\alpha/q$ increases, $V(u)$ can become negative and develop a dip on the left of the peak, which is quite different from the shape of $V(u)$ in the minimal model. For large frequencies this does not affect the reflection coefficient much, thus at large frequencies Re$(\sigma)$ behaves similarly to the minimal model. While at small frequencies,
the reflection coefficient is affected greatly and thus when $\alpha/q$ is large, Re$(\sigma)$ exhibits novel behavior compared to the minimal model.

%%%%%%%%%%%%%%%%%%%%%%%%%%%%%%%%%%%%%%
%%%%%%%%%%%%%%%%%%%%%%%%%%%%%%%%%%%%%%

%%%%%%%%%%%%%%%%%%%%%%%%%%%%%%%%%%%%%%
%%%%%%%%%%%%%%%%%%%%%%%%%%%%%%%%%%%%%%

\section{Conclusion and Discussion}

In this paper we studied a next-to-minimal realization of holographic superconductors from Einstein-Maxwell-dilaton gravity. There are two adjustable constants in our model: $\alpha$ and $q$. $\alpha$ describes the coupling between the dilaton and the Maxwell field and $q$ is the charge of the dilaton.

For the values of $\alpha$ and $q$ we considered, we found that there is always a critical temperature at which a second order phase transition occurs between a hairy black hole and the AdS RN black hole in the canonical ensemble. Below this temperature the dual theory is in a superconducting phase while above this temperature the dual theory is in a normal phase.

We calculated the value of the condensates in the superconducting phase and found that as $\alpha$ increases with fixed $q$, the value of the condensate gets smaller. We also studied the electric conductivity of the dual superconductor and found that for the values of $\alpha$ and $q$ where $\alpha/q$ is small the dual superconductor has similar properties to the minimal model as pointed out in \cite{Aprile:2009ai}. However, for the values of $\alpha$ and $q$ where $\alpha/q$ is large enough, the electric conductivity of the dual superconductor exhibits novel properties which are very different from the minimal model at low frequencies: a ``Drude Peak" arises at $\omega\to 0$ and the height of the peak depends on the temperature in a non-monotonic way. This can be seen from the shape of the potential $V(u)$ which becomes negative at some finite $u$. For all the values of $\alpha$ and $q$ we considered, the imaginary part of the electric conductivity always has a minimum at $\omega/T_c\approx 8$ for the operator ${\mathcal{O}}_2$, which is the same as in the minimal model. It is expected that $\omega_g/T_c$ for the operator ${\mathcal{O}}_2$ is also approximately $8$ for all the values of $\alpha$ and $q$.

One immediate question is about the zero-temperature limit of the superconductor as in this paper we mainly focus on the finite temperature behavior. It would be interesting to add magnetic fields to this system to study the Meissner effect of the dual superconductor and to understand the structure of fermion spectral functions in this system \cite{Chen:2009pt,{Faulkner:2009am},{Gubser:2009dt},{Basu:2010ak}}. It would also be interesting to try to embed these phenomenological models into the framework of string theory, just as in \cite{Gauntlett:2009bh}, and add DC currents to the dual field theory to study the effects of the DC currents as in \cite{Arean:2010xd}.

\section*{Acknowledgments}

We would like to thank Rong-Gen Cai and Tian-Jun Li for encouragements and supports. This work is supported in part by the Chinese Academy of Sciences with Grant No.
KJCX3-SYW-N2 and the NSFC with Grant No. 10821504 and No. 10525060.

\section*{Appendix}

In this appendix we show that the equation of motion for $g_{xx}$ can be derived from other four equations of motion.
With our ansatz (\ref{ansatzsol}) for  the solutions, there are only three nonzero equations of motion (\ref{eomgmunu}) for $g_{\mu\nu}$, {\it i.e.} the equations of motion for $g_{rr}$, $g_{tt}$ and $g_{xx}$.
These three equations of motion are 
\bea
-g^{tt}R_{tt}+g^{rr}R_{rr}+2g^{xx}R_{xx}+\frac{6}{\ell^2}U(\eta)-\frac{1}{2}\nabla^\alpha\eta\nabla_\alpha\eta
+\frac{1}{4}G(\eta)F^2+\frac{1}{2}J(\eta)A^2&=&0,\label{app1}\\
g^{tt}R_{tt}-g^{rr}R_{rr}+2g^{xx}R_{xx}+\frac{6}{\ell^2}U(\eta)+\frac{1}{2}\nabla^\alpha\eta\nabla_\alpha\eta
+\frac{1}{4}G(\eta)F^2-\frac{1}{2}J(\eta)A^2&=&0,\label{app2}\\
g^{tt}R_{tt}+g^{rr}R_{rr}+\frac{6}{\ell^2}U(\eta)-\frac{1}{2}\nabla^\alpha\eta\nabla_\alpha\eta
-\frac{1}{4}G(\eta)F^2-\frac{1}{2}J(\eta)A^2&=&0.\label{app3}
\eea
Now our task is to obtain (\ref{app3}) from (\ref{app1}) and (\ref{app2}). (\ref{app1}) and (\ref{app2}) can be simplified to be
\bea
g^{tt}R_{tt}-g^{rr}R_{rr}&=&-\frac{1}{2}\nabla^\alpha\eta\nabla_\alpha\eta
+\frac{1}{2}J(\eta)A^2,\label{app4}\\
g^{xx}R_{xx}&=&-\frac{3}{\ell^2}U(\eta)
-\frac{1}{8}G(\eta)F^2.\label{app5}
\eea
To prove the independence, we need to prove that we can obtain (\ref{app3}) from (\ref{app4}) and (\ref{app5}).

For the metric obeying the form of the ansatz (\ref{ansatzsol}), we have
\bea\label{app6} g^{tt}R_{tt}+g^{rr}R_{rr}&=&2g^{xx}R_{xx}+r(g^{xx}R_{xx})'+\frac{1}{2}(g^{tt}R_{tt}-g^{rr}R_{rr})
\bigg[1+r\frac{g'}{g}-r\chi'\bigg]\nonumber\\
&&
+\frac{1}{2}\bigg[r(g^{tt}R_{tt}-g^{rr}R_{rr})\bigg]'.\eea

 Substituting (\ref{app4}) and (\ref{app5}) into the above equality (\ref{app6}), and using (\ref{eometa})
and the following useful expression
\be\partial_r\partial^r\eta=\nabla^2\eta-(\frac{2}{r}-\frac{\chi'}{2})\partial^r\eta,\ee
we can obtain
\bea\label{app7}
g^{tt}R_{tt}+g^{rr}R_{rr}&=&-\frac{6}{\ell^2}U(\eta)+\frac{1}{2}J(\eta)A^2
+\frac{1}{2}\nabla_\mu\eta\nabla^\mu\eta-\frac{1}{4}G(\eta)F^2+
r\bigg[-\frac{1}{4}\partial_{\eta} G(\eta)\eta'F^2\nonumber\\
&&-\frac{1}{8}G(\eta)(F^2)'
+\frac{1}{4}J(\eta)(A^2)'+\frac{1}{4f}J(\eta)A^2(g'-g\chi')\bigg].
\eea

From the equation of motion for the gauge field (\ref{eomamu}), {\it i.e.}
\be \label{app8} (-\frac{\chi'}{2}+\frac{2}{r})G(\eta)F^{rt}+\partial_{\eta}G(\eta)\eta' F^{rt}+G(\eta)\partial_{r}F^{rt}=JA^{t},\ee
we obtain the following identity on shell
\be\label{app9} G(\eta)(F^2)'=-\frac{4}{r}G(\eta)F^2-2\partial_{\eta}G(\eta)
\eta'F^2+4J(\eta)F_{rt}A^{t},\ee
and at the same time, we also have the following off-shell relation
\be\label{app10}(A^2)'=2F_{rt}A^t+(\chi'-\frac{g'}{g})A^2.\ee
 Substituting (\ref{app9}) and (\ref{app10}) into (\ref{app7}), we obtain (\ref{app3}). Thus only two equations of motion in (\ref{eomgmunu}) are independent.

\end{document}